\documentclass[9pt,twocolumn,twoside]{opticajnl}

\journal{opticajournal} 

\usepackage{lineno}
\usepackage{float}
\usepackage{booktabs}
\usepackage{siunitx}
\usepackage{soul}
\sethlcolor{yellow}
\usepackage{tabularx}

\title{A heterogeneously integrated coupled-cavity frequency beam splitter}

\author[1,$\dagger$]{Lucas M. Cohen}
\author[2]{Manuel H. Mu\~{n}oz-Arias}
\author[2]{Mohan Sarovar}
\author[1]{Nicholas A. Boynton}
\author[1]{Shawn C. Arterburn}
\author[1]{Thomas A. Friedmann}
\author[1,$\ddagger$]{Nils T. Otterstrom}
\author[1,*]{Paul S. Davids}
\author[1,$\ddagger$]{Michael Gehl}

\affil[1]{Photonic and Phononic Microsystems, Sandia National Laboratories, Albuquerque, NM 87123, USA.}
\affil[2]{Quantum Algorithms and Applications Collaboratory, Sandia National Laboratories, Livermore, CA 94550, USA.}
\affil[$\dagger$]{Present address: Quantinuum, Albuquerque, NM 87113, USA.}
\affil[$\ddagger$]{Present address: Manzano Systems, Albuquerque, NM 87110, USA.}

\affil[*]{pdavids@sandia.gov}

\begin{abstract}
Frequency encoded photonic qubits promise a scalable path towards high-dimensional quantum information processing, but require efficient components for coherently mixing frequency modes. Coupled cavity modulators provide this functionality by using only a single driving microwave tone to couple hybridized optical supermodes.
Here, we demonstrate a heterogeneously integrated thin-film lithium-niobate-on-silicon coupled-cavity modulator that realizes tunable bidirectional frequency mode transformations, including \(50/50\) beam splitting and near complete frequency swapping with \(>20~\mathrm{dB}\) pump extinction at a \(10~\mathrm{GHz}\) supermode splitting. Because the electro-optic film is bonded onto a foundry fabricated silicon photonics platform, the approach is compatible with co-integration of photon pair sources, spectral filters, active tuning elements, and single photon detectors. We also bond thin-film lithium tantalate onto the same coupled-cavity platform, demonstrating material flexibility for scalable integrated frequency bin quantum photonic circuits.

\end{abstract}

\setboolean{displaycopyright}{false} 

\begin{document}

\maketitle

\section{Introduction}
Frequency encoding represents photonic quantum information using single photons in superpositions of discrete frequency modes within a common spatial channel~\cite{lukens2016frequency}. Compared to polarization~\cite{sansoni2014integrated}, spatial~\cite{politi2009integrated}, or time-bin encodings~\cite{bouchard2024programmable}, frequency encoding can support high dimensionality 
(limited by the fiber or waveguide bandwidth) within a single spatial channel and has therefore become an attractive route towards fault-tolerant quantum information processing~\cite{lu2023frequency,myilswamy2025chip}. A foundational element of frequency-encoded architectures is a deterministic, low-loss device that coherently mixes frequency modes: a frequency-mode beam splitter.

\begin{figure*}[t]
\centering
\includegraphics[width=\textwidth]{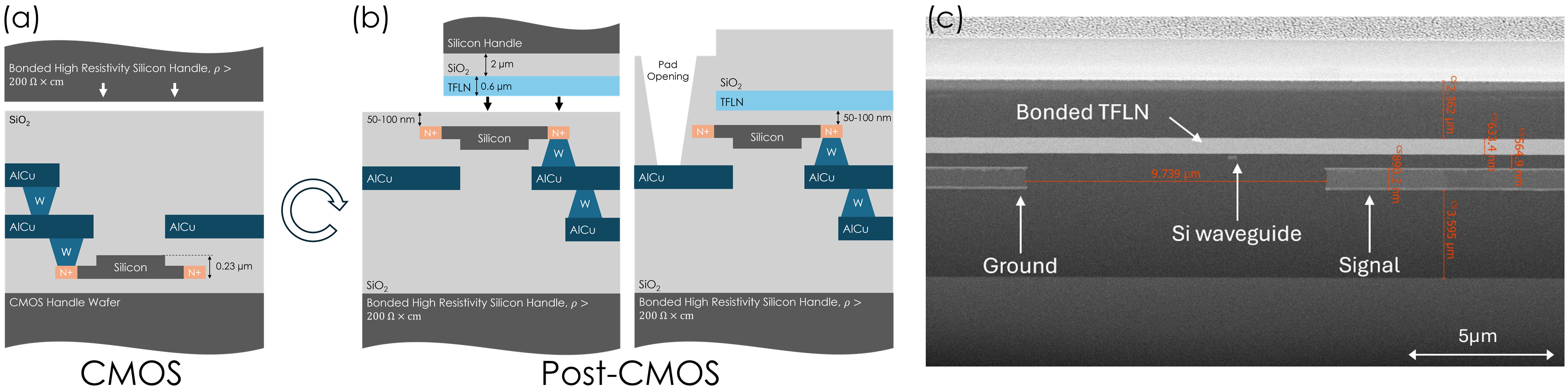}
\caption{Heterogeneous Si/TFLN platform fabrication. (a) Process flow performed in the CMOS foundry. (b) Post-foundry HI and microfabrication process. (c) SEM cross-section image of the fully processed platform.}
\label{fig1}
\end{figure*}

Coupled optical resonators, often referred to as photonic molecules, provide a general route to engineering hybridized optical supermodes and have been explored for a variety of classical and quantum photonic applications~\cite{liao2020photonic,gentry2014tunable,Wade:15,McKenna:20,gevorgyan2020active,zhang2021squeezed}. Electro-optic (EO) coupled-cavity modulators (CCMs) build on this concept by using a microwave drive to couple the hybridized supermodes of two optical resonators providing frequency-domain beam-splitter functionality~\cite{zhang2019electronically}. CCMs can perform bidirectional frequency-mode beam splitting, with the splitting ratio controlled by the microwave drive amplitude. Following early classical demonstrations of EO CCMs on monolithic thin-film lithium niobate (TFLN) platforms~\cite{zhang2019electronically,Holzgrafe:20,hu2021chip}, CCMs have recently also been demonstrated in the single-photon regime~\cite{yang2026quantum}. However, CCMs are only one component of a quantum photonic system: ideally, the same photonic chip would host entangled photon-pair sources~\cite{savanier2016photon}, spectral filters~\cite{gehl2017active}, phase shifters~\cite{watts2013adiabatic}, and single-photon detectors~\cite{martinez2017single}. Although monolithic TFLN platforms provide excellent EO performance and can support some of these functions, they do not yet offer the same breadth of mature foundry-verified components available in silicon photonics. Moreover, high-quality processing of TFLN remains challenging, and TFLN foundry offerings are still limited.

In this work, we use a mature silicon photonics (SiP) platform as the base for heterogeneous integration of TFLN by die bonding. The SiP devices are fabricated in a CMOS-compatible foundry process, after which a TFLN layer is bonded above the silicon waveguide layer to provide electro-optic functionality~\cite{boynton2020heterogeneously}. We report the design, fabrication, and characterization of a Si/TFLN CCM that performs tunable, bidirectional frequency-mode transformations, including balanced $50/50$ beam splitting and near-complete frequency swapping with greater than $20$~dB pump suppression. As a demonstration of material flexibility, we also integrate thin-film lithium tantalate (TFLT) onto the same CCM platform using the same bonding process.


\section{Theory}
\label{sec:op}

Frequency-mode beam splitters can, in principle, be implemented using EO phase modulators (EOPM). A sinusoidal RF drive modulates the optical phase and scatters input photons into sidebands spaced by the RF frequency with amplitudes that scale following Bessel functions of the modulation depth. Although EOPMs can be nearly lossless in the full frequency basis, the effective transformation is generally non-unitary when restricted to a finite frequency mode subspace. 
It has been shown~\cite{lukens2016frequency} that one can control sideband interferences via an appropriate concatenation of EOPMs and line-by-line Fourier transform pulse shapers in order to realize near-unitary frequency-mode transformations. However, in practice, such systems are complex to control and incur a significant loss penalty. The CCM device described in this work relies instead on resonant coupling between engineered supermodes by an RF phase modulation to confine the transformation primarily to a two-mode frequency subspace. The achievable splitting ratio and insertion loss are then governed by the relative rates of external coupling and intrinsic loss, rather than by suppression of a large set of modulation sidebands.

Linear input-output modeling leads to a relationship between the two supermode amplitudes of the CCM following Refs.~\cite{hu2021chip,munoz2026modeling}, and can be written as 
\begin{equation}
    \mathbf{b}_\mathrm{out}(\omega)=\Xi(\omega)\,\mathbf{b}_\mathrm{in}(\omega),
\end{equation}
with transfer matrix
\begin{equation}
\mathbf{\Xi}=
\begin{bmatrix}
1-\dfrac{2\kappa_e\kappa}{\kappa^2+\Omega^2} &
i\dfrac{2\kappa_e\Omega}{\kappa^2+\Omega^2}e^{i\phi}
\\[1.1em]
i\dfrac{2\kappa_e\Omega}{\kappa^2+\Omega^2}e^{-i\phi} &
1-\dfrac{2\kappa_e\kappa}{\kappa^2+\Omega^2}
\end{bmatrix},
\label{eq2}
\end{equation}
with \(\kappa_e\) and \(\kappa\) denoting the external bus-coupling rate and total loss rate, and $\Omega$ and $\phi$ are the RF modulation amplitude and phase, respectively. The output power in either supermode is influenced by $\Omega$ together with the dimensionless parameter 
\begin{equation}
\alpha \equiv \frac{\gamma}{\kappa_i} = \frac{2\kappa_e}{\kappa_i} = \frac{Q_i}{Q_e},
\label{eq:alpha}
\end{equation}
the ratio of intrinsic ($\kappa_i$) to extrinsic (coupling, defined as $\gamma = 2 \kappa_e$) losses, which governs the loss of the device as well as the operating regime. Under this definition, $\alpha>1$ corresponds to the overcoupled regime, $\alpha=1$ to critical coupling, and $\alpha<1$ to the undercoupled regime.

We highlight two important drive amplitudes. First, at $\Omega_{BS}^{\pm}=\lvert\gamma\pm\sqrt{2\gamma^2-\kappa_i^2}\rvert$, the transfer matrix becomes
\begin{equation}
    \mathbf{\Xi}_{BS}=\frac{\sqrt{2}\pm\sqrt{\alpha^2-2}}{\alpha+2}\,\mathbf{K},
    \qquad
    \mathbf{K}=\frac{1}{\sqrt{2}}
    \begin{bmatrix} 1 & ie^{i\phi}\\ ie^{-i\phi} & 1 \end{bmatrix},
\end{equation}
which is a $50/50$ balanced frequency beam splitter with a loss determined by $\alpha$ and that exists only when $\alpha>\sqrt{2}$. Lastly, at the generalized critical coupling (GCC) amplitude $\Omega_{GCC}=\sqrt{\gamma^2-\kappa_i^2}$ the diagonal terms vanish and 
\begin{equation}
    \mathbf{\Xi}_{GCC}=
    \begin{bmatrix}
        0 & i\sqrt{\dfrac{\alpha-2}{\alpha+2}}\,e^{i\phi}\\[3mm]
        i\sqrt{\dfrac{\alpha-2}{\alpha+2}}\,e^{-i\phi} & 0
    \end{bmatrix},
\end{equation}
which is a $100/0$ frequency shifter again with loss determined by $\alpha$ and existing only when $\alpha>2$. Sweeping the drive amplitude $\Omega$ across the ranges shown here continuously tunes the splitting ratio, while $\alpha$ fixes the insertion loss for a given device.

\begin{figure*}[t]
\centering
\includegraphics[width=\textwidth]{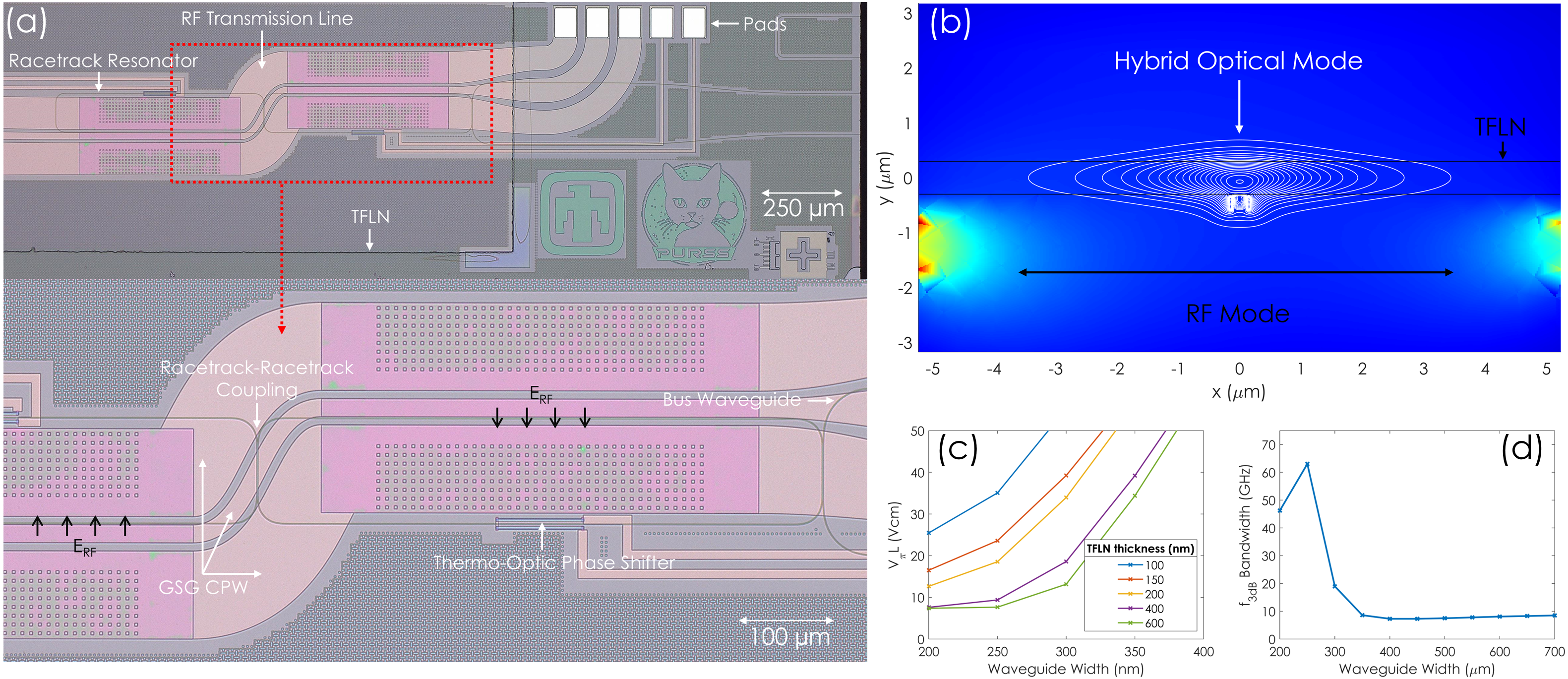}
\caption{CCM device design. (a) Optical microscope image of the CCM modulator. (b) Simulated overlay of the $E_{x}$ component of the optical and RF fields in the EO modulation region. (c) Simulated single-arm $V_{\pi}L$ as a function of TFLN slab thickness and silicon waveguide width. (d) Estimated RF $3$~dB bandwidth for a \SI{250}{nm}-wide silicon waveguide with \SI{600}{nm}-thick TFLN layer.}
\label{fig2}
\end{figure*}

\section{Fabrication}

A cross section of the HI platform fabricated via Sandia’s Microsystems Engineering, Science and Applications (MESA) complex is shown in Fig.~\ref{fig1}(a-b).  Fabrication begins on a silicon-on-insulator (SOI) wafer with a \SI{200}{nm} buried oxide layer and a \SI{270}{nm} device layer.  After oxidation, which reduces the thickness of the device layer to \SI{230}{nm}, both strip and rib type waveguides are formed through a combination of two reactive ion etching steps (a partial \SI{80}{nm} etch, followed by a full thickness etch).  Resistive features for thermal phase shifters are formed though an n+ type dopant.  Tungsten vias provide electrical contact to these thermal phase shifters.  Two metal layers, connected via Tungsten vias, are used for DC and RF electrical signals.  The lower metal layer is ultimately in closer proximity to the EO material and is therefore used for RF co-planar waveguides.  The upper layer is further from the optical waveguide and is therefore primarily used for routing signals over waveguides with minimal loss.  The metal layers are encapsulated with an upper oxide layer which is carefully planarized.  This entire wafer is then flipped and bonded onto a new handle wafer, as illustrated in Fig.~\ref{fig1}(b).  The original handle is removed through a combination of grinding and TMAH-like wet etching, resulting in the buried oxide of the original SOI wafer now being exposed at the top of the photonic stack.  This buried oxide layer is further thinned through a timed wet etch to a final thickness in the range of $50$-\SI{100}{nm}.  The wafer is then diced into individual reticles (not yet into individual die), and processing continues at the reticle level.  Chiplets of commercially sourced TFLN/TFLT (\SI{600}{nm} X-cut LiNbO$_3$, \SI{600}{nm} X-cut LiTaO$_3$, NanoLN) are bonded to each die within the reticle.  The bond is strengthened by annealing at $250^\circ$C for $2$~hrs under $3$~kN of applied force. Removal of the handle on the TFLN chips is accomplished with a similar backgrind and wet etch approach leaving the original buried oxide exposed on the LN. A \SI{2}{um}-thick CVD SiO$_2$ is then grown atop the entire sample, followed by a patterned pad etch to reveal the aluminum electrical pads to the active components on the chip.  Finally, the reticle is diced into individual die for final testing. A cross-sectional SEM image of a fully processed chip is shown in Fig.~\ref{fig1}(c).

\section{Device Design}
\label{sec:dd}

The CCM device is designed on the HI platform and consists of two evanescently coupled racetrack resonators coupled to a single bus waveguide. We employ strip silicon waveguides with a nominal width and thickness of \SI{500}{\nano\meter} and \SI{230}{\nano\meter}, respectively, which keeps the optical energy highly confined in the silicon. The resonators use linear directional couplers for bus and inter-resonator coupling, and Euler bends to form the racetrack cavities. Straight sections of \SI{575}{\micro\meter} length connect the couplers. A coplanar-waveguide (CPW) RF electrode of \SI{100}{}/\SI{20}{}/\SI{100}{\micro\meter}--wide ground--signal--ground (GSG) electrodes with a \SI{10}{\micro\meter} ground--signal gap is shared between the two resonators and arranged to drive each resonator out of phase (Fig.~\ref{fig2}(a)).  

The EO modulation is applied on one side of the straight sections of the racetracks. Here, the waveguide width tapers linearly over \SI{125}{\micro\meter} to \SI{250}{\nano\meter}, resulting in a hybrid mode spanning both the silicon and TFLN with $\sim 65\%$ of the optical energy in the TFLN (The $E_{x}$ field component of the hybrid optical mode and the CPW RF mode in this region are depicted in white contours and colorbar, respectively, in Fig.~\ref{fig2}(b)). Using a \SI{250}{\nano\meter}--wide silicon waveguide with a \SI{600}{\nano\meter}--thick TFLN film with \SI{50}{nm}-thick silicon dioxide interlayer, simulations predict a single-arm half-wave voltage-length product of $V_{\pi}L \sim \SI{7.6}{\volt\cm}$ and modulation bandwidths $> 50$~GHz with a taper insertion loss $< 10\%$. Device performance is strongly influenced by interlayer oxide thickness: a thicker layer increases taper loss and $V_{\pi}$, whereas a thinner layer generally improves performance. Outside of the modulation regions the RF field is transferred to a lower metal layer via Tungsten vias to avoid free-carrier induced optical loss when crossing the waveguide. Doped silicon slabs adjacent to each resonator function as thermo-optic phase shifters for static DC tuning of the individual resonance frequencies.

\begin{figure}[!htb]
\centering\includegraphics[width=\linewidth]{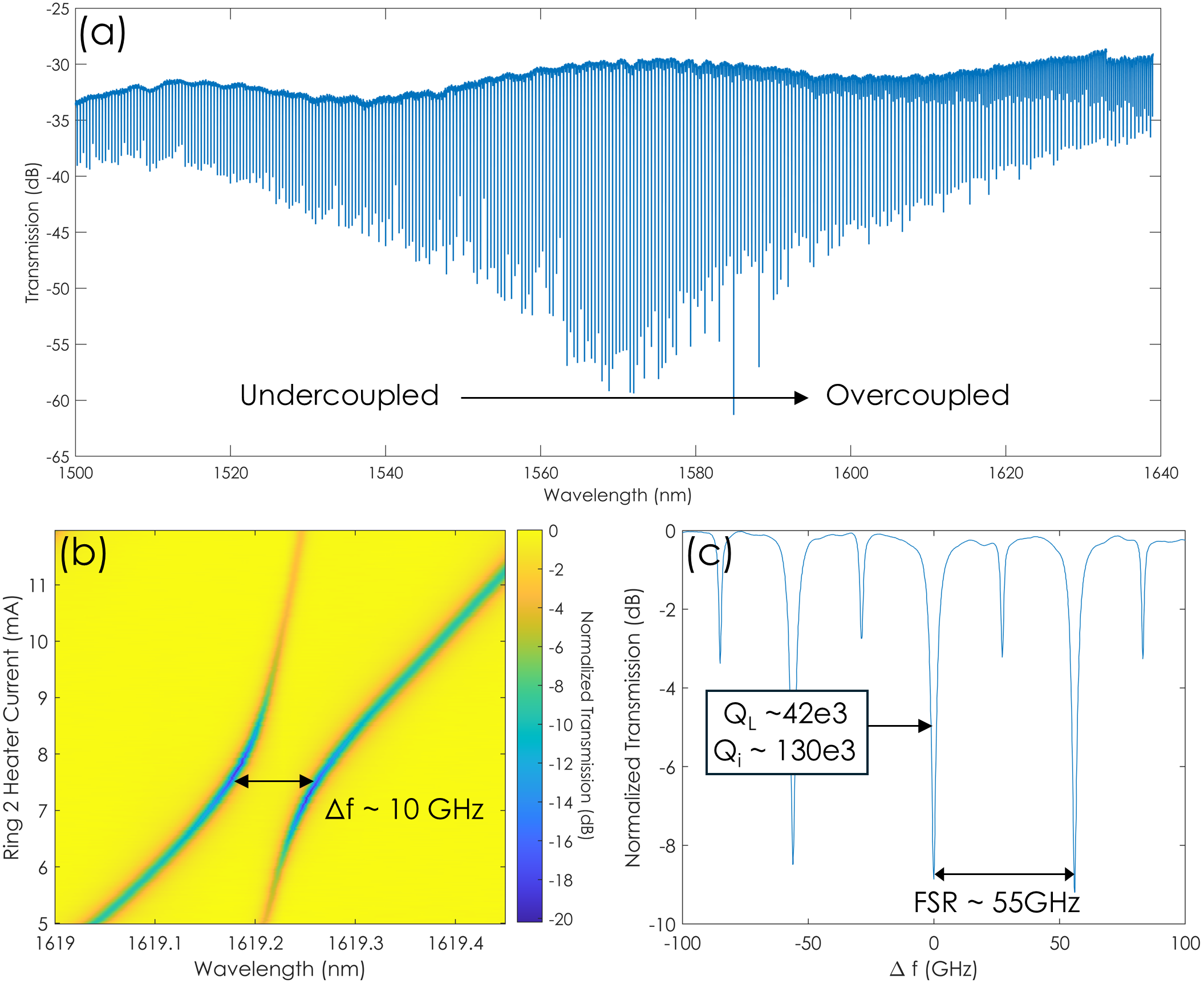}
\caption{Passive optical characterization of the CCM. (a) Broadband transmission spectrum of the untuned device. (b) Thermally tuned avoided mode crossing near \SI{1620}{nm} center wavelength, showing a supermode splitting $\sim 10$~GHz. (c) Zoomed-in transmissions spectrum with resonators in the CCM detuned, used to extract quality factors.}
\label{fig3}
\end{figure}

\section{Linear Characterization}

We first characterize the passive optical response of the device. The directional couplers used exhibit a strong spectral dependence producing the wavelength dependent extinction seen in Fig.~\ref{fig3}(a). 
Because the CCM performs better with large $\alpha$ (as discussed in Sec \ref{sec:op}), we operate near \SI{1620}{\nm} in the overcoupled regime. We tune the resonators across a common resonance mode at \SI{1620}{\nm} using the thermo-optic phase shifters and measure an avoided mode crossing with a supermode splitting of $\sim 10$~GHz in Fig.~\ref{fig3}(b). We fit the spectrum when resonators are far detuned and find a single-resonator free spectral range (FSR) of $\sim 55$~GHz, loaded quality factor of $Q_L \sim 42 \times 10^3$, and an intrinsic quality factor of $Q_i \sim 130 \times 10^3$ ($Q_e \sim 62 \times 10^3$). These values place the device in the overcoupled regime with $\alpha \sim 2.23$ following Eq.~\ref{eq:alpha}. 

\section{Radio-Frequency Operation}

For RF measurements, we align a tunable continuous-wave telecom laser to one optical supermode of the CCM and drive the CPW electrode at the measured supermode splitting frequency. The RF drive power is swept while the output is split and monitored simultaneously by an optical spectrum analyzer (OSA) and a power meter. The OSA records the distribution of optical power between the pumped and converted supermodes, while the power meter records the total transmitted optical power. Optical powers are normalized to the transmission of the CCM when pumped off-resonance. RF powers correspond to the calibrated power after amplification and cable losses at the input electrical probe.

\begin{figure}[!htb]
\centering\includegraphics[width=\linewidth]{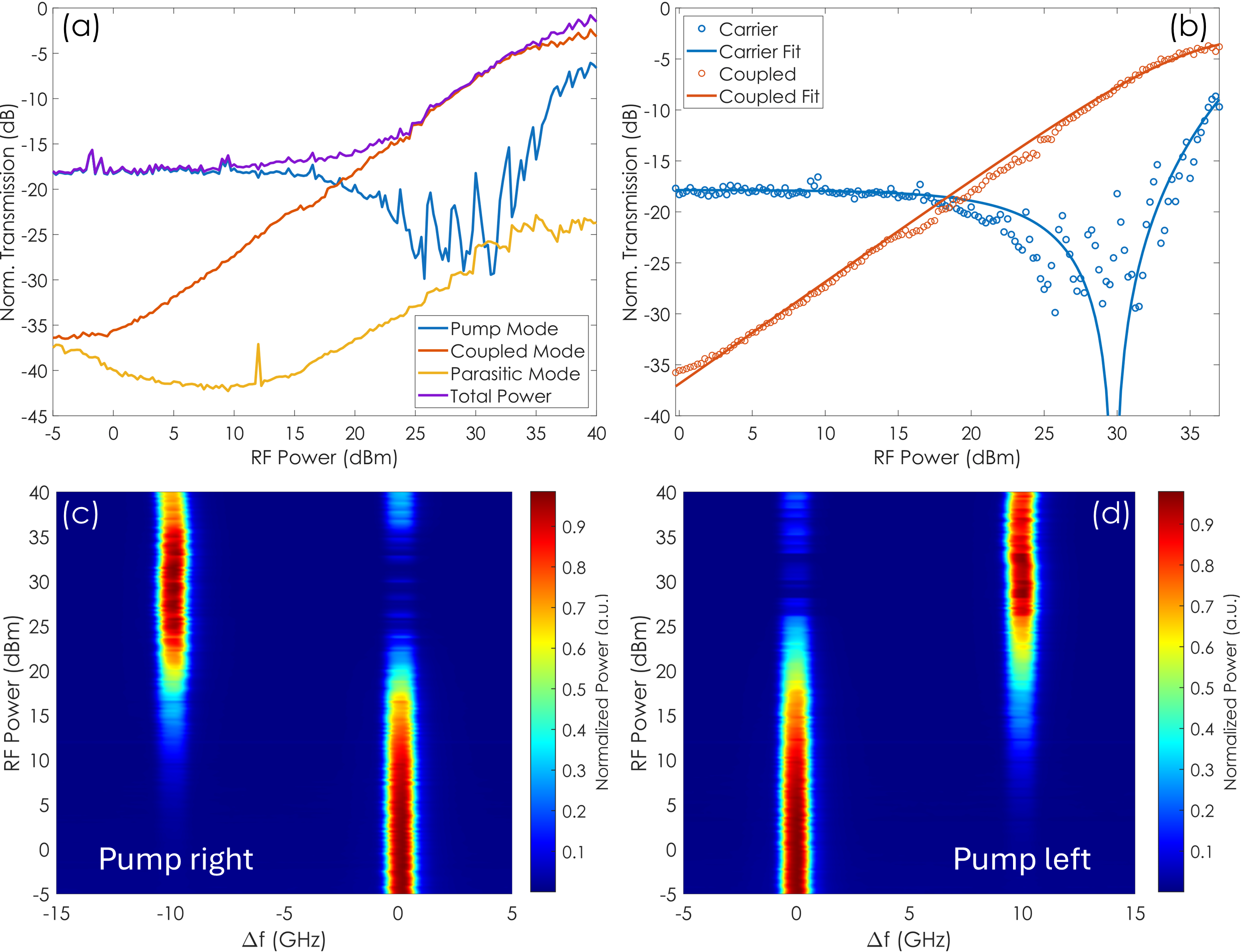}
\caption{RF-driven conversion in the Si/TFLN CCM. (a) Measured conversion efficiency versus RF drive power while pumping the high-frequency supermode. (b) Measured data and fit conversion efficiencies. c,d) Conversion efficiency spectrogram while pumping the high- (c) and low- (d) frequency optical supermode.}
\label{fig4}
\end{figure}

The results measured while pumping the high-frequency supermode are shown in Fig.~\ref{fig4}(a). At low RF drive, optical energy stays in the pump supermode with very little conversion to the partner supermode; the measured $\sim -17$~dB transmission corresponds to the zero-RF extinction of the CCM device, as shown in Fig.~\ref{fig4}(b)). As the RF drive is increased, power converted into the partner supermode initially increases approximately linearly with RF drive power, while the optical power in the pump supermode is reduced. In addition, the total transmitted power increases with RF drive power. This reduction in apparent loss and behavior is the EO analogue of Autler--Townes splitting in a driven photonic molecule, as discussed in Ref.~\cite{zhang2019electronically}. At $\sim 18.5$~dBm of RF drive power, the device acts as a $50/50$ frequency beam-splitter between the two supermodes, and at $\sim 30$~dBm drive power the optical energy is nearly entirely transferred to the partner supermode with $> 20$~dB extinction of the pump. Increasing the RF power beyond this point transfers power back from the partner supermode towards the pumped supermode. We measure $> 15$~dB of suppression of parasitic sidebands (frequencies outside the two-frequency-supermode subspace) for all drive powers. Finally, Figs.~\ref{fig4}(c,d) show the measured RF/optical spectrograms while pumping the high- and low-frequency supermodes of the CCM, respectively. At each RF frequency, the power spectral density is normalized to the total optical power. The two cases exhibit very similar performance, as expected from theory.

We proceed to fit the measured frequency conversion spectra of Fig.~\ref{fig4}(a) to the transfer matrix of Equation~\ref{eq2} in the linear regime of the RF amplifier ($0 - 35$~dBm output power). The resulting fit is plotted with the data in Fig.~\ref{fig4}(b), from which we extract $\kappa_e/2\pi \sim$~\SI{1.85}{GHz}, $\kappa_i/2\pi \sim$~\SI{1.51}{GHz}, and $\alpha \sim 2.45$, in good agreement with the values obtained independently from linear transmission spectrum fitting ($\alpha \sim 2.23$). The corresponding half-wave voltage--length product is $V_{\pi}L \sim \SI{8.28}{\volt\cm}$ for a single (non-push--pull) electro-optic arm with a $\sim$~\SI{325}{\micro\meter}--long phase shifter. The measured value is consistent with that expected from simulation ($\sim$~\SI{7.6}{\volt\cm}) when considering the strong dependence on performance with the Si-TFLN interlayer oxide thickness. The extracted parameters are tabulated in the first row of Table~\ref{tab:table1}.

\begin{figure}[H]
\centering\includegraphics[width=\linewidth]{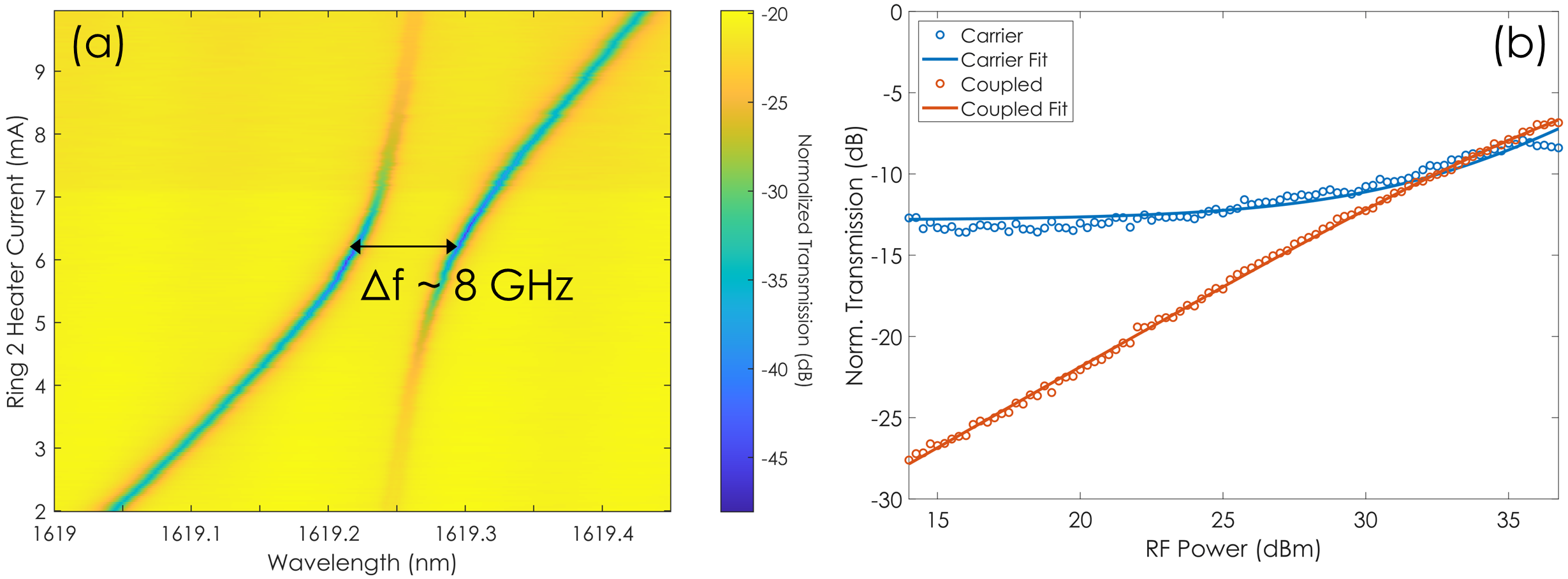}
\caption{Material flexibility demonstration using bonded TFLT. (a) Avoided mode crossing for the CCM with bonded TFLT. (b) Measured and fitted RF conversion efficiencies.}
\label{fig5}
\end{figure}

As a demonstration of the material flexibility enabled by this HI approach, we bond \SI{600}{nm}-thick die of TFLT atop our CCM device. The TFLT devices show weaker bus-waveguide and resonator-resonator couplings, which we attribute to the lower refractive index of TFLT, and a slightly increased optical loss in the hybrid waveguide tapers. The measured avoided mode crossing and frequency conversion is shown in Fig.~\ref{fig5}(a-b). From the fitted spectra, we estimate a $V_{\pi}L \sim$~\SI{9}{\volt \cm} with additional metrics tabulated in Table~\ref{tab:table1}. This larger $V_\pi L$ should not be taken as conclusive evidence of inferior TFLT material performance, since the modulation efficiency is strongly dependent on the interlayer oxide thickness, as discussed in Sec.~\ref{sec:dd}. The extracted $\alpha \sim 1.46$ is sufficient for balanced beam splitting in this model but below the threshold for GCC, consistent with the reduced conversion efficiency observed experimentally.

\begin{table}[H]
    \centering
    \begin{tabularx}{\linewidth}{lXXXXXX}
    \toprule
         Film & Splitting (GHz) &  $\kappa_e/2\pi$ (GHz) & $\kappa_i/2\pi$ (GHz) & $\alpha$ & $V_{\pi}$ (V) & $V_{\pi}L$ (V\,cm) \\
     \midrule
         TFLN & 10 & 1.85 & 1.51 & 2.45 & 255 & 8.28 \\
         TFLT & 8 & 1.22 & 1.67 & 1.46 & 275 & 9 \\ 
     \bottomrule
    \end{tabularx}
    \caption{Electro-optic fit parameters extracted from fitting the measured frequency conversion spectra with Eq.~\ref{eq2}.}
    \label{tab:table1}
\end{table}

\section{Conclusion}

We have demonstrated a heterogeneously integrated Si/TFLN coupled-cavity frequency beam splitter. By varying the RF drive amplitude, the device is tuned continuously from weak conversion through balanced 50/50 frequency beam splitting to near-complete frequency swapping with greater than 20 dB suppression of the pumped supermode. Because the electro-optic layer is integrated after foundry SiP fabrication, this approach is compatible with co-integration of photon-pair sources, spectral filters, active switches, multiplexers, and single-photon detectors. These results provide a route toward scalable, fully integrated frequency-bin quantum photonic circuits. Ongoing work targets reduced optical loss through silicon-nitride-based platforms and integration of CCMs with additional components toward fully integrated quantum circuits.

\section*{Acknowledgments}
We acknowledge support from the Sandia National Laboratories Laboratory Directed Research and Development program (EPIQ project). Sandia National Laboratories is a multimission laboratory managed and operated by National Technology and Engineering Solutions of Sandia LLC, a wholly owned subsidiary of Honeywell International Inc. for the U.S. Department of Energy’s National Nuclear Security Administration under contract DE-NA0003525.

\bibliography{sample}

@article{boynton2020heterogeneously,
  title={A heterogeneously integrated silicon photonic/lithium niobate travelling wave electro-optic modulator},
  author={Boynton, Nicholas and Cai, Hong and Gehl, Michael and Arterburn, Shawn and Dallo, Christina and Pomerene, Andrew and Starbuck, Andrew and Hood, Dana and Trotter, Douglas C and Friedmann, Thomas and others},
  journal={Optics express},
  volume={28},
  number={2},
  pages={1868--1884},
  year={2020},
  publisher={Optical Society of America}
}

@article{lu2023frequency,
  title={Frequency-bin photonic quantum information},
  author={Lu, Hsuan-Hao and Liscidini, Marco and Gaeta, Alexander L and Weiner, Andrew M and Lukens, Joseph M},
  journal={Optica},
  volume={10},
  number={12},
  pages={1655--1671},
  year={2023},
  publisher={Optica Publishing Group}
}

@article{myilswamy2025chip,
  title={On-chip frequency-bin quantum photonics},
  author={Myilswamy, Karthik V and Cohen, Lucas M and Seshadri, Suparna and Lu, Hsuan-Hao and Lukens, Joseph M},
  journal={Nanophotonics},
  volume={14},
  number={11},
  pages={1879--1894},
  year={2025},
  publisher={De Gruyter}
}

@article{hu2021chip,
  title={On-chip electro-optic frequency shifters and beam splitters},
  author={Hu, Yaowen and Yu, Mengjie and Zhu, Di and Sinclair, Neil and Shams-Ansari, Amirhassan and Shao, Linbo and Holzgrafe, Jeffrey and Puma, Eric and Zhang, Mian and Lon{\v{c}}ar, Marko},
  journal={Nature},
  volume={599},
  number={7886},
  pages={587--593},
  year={2021},
  publisher={Nature Publishing Group UK London}
}

@article{zhang2019electronically,
  title={Electronically programmable photonic molecule},
  author={Zhang, Mian and Wang, Cheng and Hu, Yaowen and Shams-Ansari, Amirhassan and Ren, Tianhao and Fan, Shanhui and Lon{\v{c}}ar, Marko},
  journal={Nature Photonics},
  volume={13},
  number={1},
  pages={36--40},
  year={2019},
  publisher={Nature Publishing Group UK London}
}

@article{lukens2016frequency,
  title={Frequency-encoded photonic qubits for scalable quantum information processing},
  author={Lukens, Joseph M and Lougovski, Pavel},
  journal={Optica},
  volume={4},
  number={1},
  pages={8--16},
  year={2016},
  publisher={Optical Society of America}
}

@article{yang2026quantum,
  title={Quantum photonic frequency processor on thin-film lithium niobate},
  author={Yang, Ran and Zhou, Wei and Guo, Dong-Jie and Ke, Hong-Ming and Tao, Linrunde and Wei, Ying and Duan, Jia-Chen and Cui, Yu and Jia, Kunpeng and Xie, Zhenda and others},
  journal={arXiv preprint arXiv:2603.11471},
  year={2026}
}

@article{savanier2016photon,
  title={Photon pair generation from compact silicon microring resonators using microwatt-level pump powers},
  author={Savanier, Marc and Kumar, Ranjeet and Mookherjea, Shayan},
  journal={Optics express},
  volume={24},
  number={4},
  pages={3313--3328},
  year={2016},
  publisher={Optical Society of America}
}

@article{gehl2017active,
  title={Active phase correction of high resolution silicon photonic arrayed waveguide gratings},
  author={Gehl, M and Trotter, D and Starbuck, A and Pomerene, A and Lentine, AL and DeRose, C},
  journal={Optics Express},
  volume={25},
  number={6},
  pages={6320--6334},
  year={2017},
  publisher={Optical Society of America}
}

@article{watts2013adiabatic,
  title={Adiabatic thermo-optic Mach--Zehnder switch},
  author={Watts, Michael R and Sun, Jie and DeRose, Christopher and Trotter, Douglas C and Young, Ralph W and Nielson, Gregory N},
  journal={Optics letters},
  volume={38},
  number={5},
  pages={733--735},
  year={2013},
  publisher={Optical Society of America}
}

@article{martinez2017single,
  title={Single photon detection in a waveguide-coupled Ge-on-Si lateral avalanche photodiode},
  author={Martinez, Nicholas JD and Gehl, Michael and Derose, Christopher T and Starbuck, Andrew L and Pomerene, Andrew T and Lentine, Anthony L and Trotter, Douglas C and Davids, Paul S},
  journal={Optics express},
  volume={25},
  number={14},
  pages={16130--16139},
  year={2017},
  publisher={Optical Society of America}
}

@article{munoz2026modeling,
  title={Modeling integrated frequency shifters and beam splitters},
  author={Mu{\~n}oz-Arias, Manuel H and Randles, Kevin J and Otterstrom, Nils T and Davids, Paul S and Gehl, Michael and Sarovar, Mohan},
  journal={arXiv preprint arXiv:2602.06003},
  year={2026}
}

@book{sansoni2014integrated,
  title={Integrated devices for quantum information with polarization encoded qubits},
  author={Sansoni, Linda},
  year={2014},
  publisher={Springer}
}

@article{politi2009integrated,
  title={Integrated quantum photonics},
  author={Politi, Alberto and Matthews, Jonathan CF and Thompson, Mark G and O'Brien, Jeremy L},
  journal={IEEE Journal of Selected Topics in Quantum Electronics},
  volume={15},
  number={6},
  pages={1673--1684},
  year={2009},
  publisher={IEEE}
}

@article{bouchard2024programmable,
  title={Programmable photonic quantum circuits with ultrafast time-bin encoding},
  author={Bouchard, Fr{\'e}d{\'e}ric and Fenwick, Kate and Bonsma-Fisher, Kent and England, Duncan and Bustard, Philip J and Heshami, Khabat and Sussman, Benjamin},
  journal={Physical Review Letters},
  volume={133},
  number={9},
  pages={090601},
  year={2024},
  publisher={APS}
}

@article{liao2020photonic,
  title={Photonic molecule quantum optics},
  author={Liao, Kun and Hu, Xiaoyong and Gan, Tianyi and Liu, Qihang and Wu, Zhenlin and Fan, Chongxiao and Feng, Xilin and Lu, Cuicui and Liu, Yong-chun and Gong, Qihuang},
  journal={Advances in Optics and Photonics},
  volume={12},
  number={1},
  pages={60--134},
  year={2020},
  publisher={Optical Society of America}
}

@article{gevorgyan2020active,
  title={Active-cavity photonic molecule optical data wavelength converter for silicon photonics platforms},
  author={Gevorgyan, Hayk and Khilo, Anatol and Popovi{\'c}, Milo{\v{s}} A},
  journal={arXiv preprint arXiv:2005.04989},
  year={2020}
}

@article{gentry2014tunable,
  title = {Tunable coupled-mode dispersion compensation and its application to on-chip resonant four-wave mixing},
  author = {Gentry, Cale M. and Zeng, Xiaoge and Popovi{\'c}, Milo{\v{s}} A.},
  journal = {Optics Letters},
  volume = {39},
  number = {19},
  pages = {5689--5692},
  year = {2014},
  publisher = {Optica Publishing Group}
}

@article{zhang2021squeezed,
  title={Squeezed light from a nanophotonic molecule},
  author={Zhang, Yue and Menotti, M and Tan, K and Vaidya, VD and Mahler, DH and Helt, LG and Zatti, L and Liscidini, M and Morrison, B and Vernon, Z},
  journal={Nature communications},
  volume={12},
  number={1},
  pages={2233},
  year={2021},
  publisher={Nature Publishing Group UK London}
}

@article{McKenna:20,
author = {Timothy P. McKenna and Jeremy D. Witmer and Rishi N. Patel and Wentao Jiang and Rapha\"{e}l Van Laer and Patricio Arrangoiz-Arriola and E. Alex Wollack and Jason F. Herrmann and Amir H. Safavi-Naeini},
journal = {Optica},
number = {12},
pages = {1737--1745},
publisher = {Optica Publishing Group},
title = {Cryogenic microwave-to-optical conversion using a triply resonant lithium-niobate-on-sapphire transducer},
volume = {7},
month = {Dec},
year = {2020},
url = {https://opg.optica.org/optica/abstract.cfm?URI=optica-7-12-1737},
doi = {10.1364/OPTICA.397235},
}

@article{Holzgrafe:20,
author = {Jeffrey Holzgrafe and Neil Sinclair and Di Zhu and Amirhassan Shams-Ansari and Marco Colangelo and Yaowen Hu and Mian Zhang and Karl K. Berggren and Marko Lon\v{c}ar},
journal = {Optica},
number = {12},
pages = {1714--1720},
publisher = {Optica Publishing Group},
title = {Cavity electro-optics in thin-film lithium niobate for efficient microwave-to-optical transduction},
volume = {7},
month = {Dec},
year = {2020},
url = {https://opg.optica.org/optica/abstract.cfm?URI=optica-7-12-1714},
doi = {10.1364/OPTICA.397513},
}

@article{Wade:15,
author = {Mark T. Wade and Xiaoge Zeng and Milo\v{s} A. Popovi\'{c}},
journal = {Opt. Lett.},
number = {1},
pages = {107--110},
publisher = {Optica Publishing Group},
title = {Wavelength conversion in modulated coupled-resonator systems and their design via an equivalent linear filter representation},
volume = {40},
month = {Jan},
year = {2015},
url = {https://opg.optica.org/ol/abstract.cfm?URI=ol-40-1-107},
doi = {10.1364/OL.40.000107},
}

\end{document}